\documentstyle[12pt]{article}
\newlength{\extraspace}
\newlength{\extraspaces}
\newcommand{\be}{\begin{equation}
\addtolength{\abovedisplayskip}{\extraspaces}
\addtolength{\belowdisplayskip}{\extraspaces}
\addtolength{\abovedisplayshortskip}{\extraspace}
\addtolength{\belowdisplayshortskip}{\extraspace}}
\newcommand{\ee}{\end{equation}}
\newcommand{\ba}{\begin{eqnarray}
\addtolength{\abovedisplayskip}{\extraspaces}
\addtolength{\belowdisplayskip}{\extraspaces}
\addtolength{\abovedisplayshortskip}{\extraspace}
\addtolength{\belowdisplayshortskip}{\extraspace}}
\newcommand{\ea}{\end{eqnarray}}

\newcommand{\nonu}{\nonumber \\[.5mm]}
\newcommand{\A}{&\!\!\!}
\begin{document}
\thispagestyle{empty}
\setlength{\baselineskip}{6mm}
\begin{center}
{\large \bf Nonlinear-supersymmetric General Relativity Theory  I}
\\[20mm]
{\sc Kazunari Shima}
\footnote{
\tt e-mail: shima@sit.ac.jp  
}   
\\[2mm]
{\it Laboratory of Physics, 
Saitama Institute of Technology \\
Fukaya, Saitama 369-0293, Japan} \\[50mm]
\begin{abstract}
The geometrical argument of the general relativity principle of Einstein is formulated in unstable Riemann space-time just inspired by the  nonlinear representation of supersymmetry, which produces new Einstein-Hilbert type action.
They  show  a new paradigm for the supersymmetric unification of space-time
 and matter, which gives new insight into the unsolved problems of
particle physics and cosmology.      \\[5mm]

\noindent
PACS: 02.40.Ky, 04,65,+e, 11.30.Pb, 12.60.Jv, 12.60.Rc, 11.30.Rd,  \\
\noindent
Keywords: linear/nonlinear supersymmetry, Nambu-Goldstone fermion, unification of space-time and matter, Einstein-Hilbert type action 
\end{abstract}
\end{center}
\newpage
\section{ Introduction }
The symmetry and its spontaneous breaking are key notions for describing  the rationale of being of nature.
Supersymmetry (SUSY)\cite{WZ1,VA} 
related naturally to space-time symmetry is promissing for the unification of general relativity and the low enegy SM in {\it one} irreducible representation of the symmetry group. 
Therefore, the evidences of SUSY and its spontaneous breakdown \cite{SS,FI,O} should be studied  
not only in (low energy) particle physics but also in cosmology, 
i.e. in the framework necessarily  including graviton. 
SUGRA is the most promissing framework  and 
the particle assignment of SO(8) SUGRA is studied  
within the existing field theoretical models\cite{MG}. 
And  we have found by group theoretical arguments that 
among all $SO(N)$ super-Poincar\'e (sP) groups  the $SO(10)$ sP group decomposed as 
${N = {10} = {5}+{5^{*}}}$ under $SO(10) \supset SU(5)$
may be a unique and minimal group which accomodates all observed particles including graviton in {\it a single} irreducible representation of  $N=10$ {\it linear(L)} SUSY \cite{KS1}, 
In this case  
10 supercharges $Q^{I}, (I=1,2, \cdots. 10)$ are decomposed as follows:  %
{$\underbar{10}_{SO(10)}=\underbar{5}_{SU(5)}+{\underbar{5}^{*}}_{SU(5)}$},    
\begin{eqnarray}
{\underbar5^{*}}_{SU(5)}=[ \{ Q_a(a=1,2,3): \underbar3^{*c}, \underbar1^{ew}; 
({e \over 3},{e \over 3},{e \over 3}) \} + \{Q_m(m=4,5): \underbar1^{c}, {\underbar2}^{ew} (-e, 0)  \} ].
\label{su(5)}
\end{eqnarray}  
The quintet  of the superchage 
$\underbar5_{SU(5)}$ have the same quantum numbers as ${\underbar5}$ of SU(5) GUT, i.e.,  { [$\bar d$-type supercharge $Q_{a}$, \ $(e., \nu_{e})$-type  supercharges $Q_{m}$]},   
Applying the representation theory of sP algebra the massless helicity 
state $|{h}>$  of {the gravity supermultiplet}  of SO(10) sP 
is specified by  $|{h}>=Q^{n} Q^{{n-1}}\cdots Q^{2}Q^{1}| {{} 2}>,\ Q^{n}\
(n=0,1, 2,\cdots ,10)$,  for the helicity $h=( 2-{n \over 2})$. 
Interestingly,  $Q^{n} Q^{{n-1}}\cdots Q^{2}Q^{1}, n=12,\cdots,n$ is the all possible non-trivial combination(product) of the supercharge with   
the dimension  ${ d}_{[n]}={{{10!} \over {n!(10-n)!}}}$ 
(with CPT conjugation)  and produces the following massless gravity  supermultiplet.   

\begin{table}[htb]
\begin{center}
\begin{tabular}{|c||c|c|c|c|c|c|c|} \hline
\hspace{0mm} $|h|$ \hspace{0mm} & \hspace{0mm} $3$ \hspace{0mm} & 
\hspace{0mm} ${5 \over 2}$ \hspace{0mm} & 
\hspace{0mm} ${} 2$ \hspace{0mm} & \hspace{0mm} ${3 \over 2}$ \hspace{0mm} & 
\hspace{0mm} ${1}$ \hspace{0mm} & \hspace{0mm} ${1 \over 2}$ \hspace{0mm} & 
\hspace{0mm} ${0}$ \hspace{0mm}   \\ \hline 
\raise2ex\hbox{$
\begin{array}{c}
{{} {} }    \\[2mm]
{ {d}_{[n]} }  \\
\end{array} 
$}  
& 
\raise2ex\hbox{$
\begin{array}{c}
{{} {} }    \\[2mm]
{ {1}_{[10]} }  \\
\end{array} 
$}  
& 
\raise2ex\hbox{$
\begin{array}{c}
{} \\[2mm]
{ {10}_{[9]}} \\
\end{array} 
$} 
&
\raise2ex\hbox{$
\begin{array}{c}
{ {{} 1}_{[0]} }    \\[2mm]
{ {45}_{[8]} }  \\
\end{array} 
$} 
& \raise2ex\hbox{$
\begin{array}{c}
{ {10}_{[1]}}    \\[2mm]
{ {120}_{[7]} }  \\
\end{array} 
$} 
& \raise2ex\hbox{$
\begin{array}{c}
{ {45}_{[2]}}    \\[2mm]
{ {210}_{[6]}}  \\
\end{array} 
$} 
& \raise2ex\hbox{$
\begin{array}{c}
{ {120}_{[3]}}    \\[2mm]
{ {252}_{[5]}}  \\
\end{array} 
$} 
& \raise2ex\hbox{$
\begin{array}{c}
{ {210}_{[4]}}    \\[2mm]
{ {210}_{[4]}}  \\
\end{array} 
$} 
\\ \hline
\end{tabular}
\end{center}
\end{table}
%
%
In order to extract any possible low energy physical contents from the tower of the helicity states we assume a {\it maximal} $SU(3) \times SU(2) \times U(1)$ invariant superHiggs-like mass generation  mechanism among helicity states, 
i.e., all  high helicity  states   redundant for SM become massive by absorbing  lower helicity states as the longitudinal components  in SM invariant way. 
Many lower helicity states disappear fromthe physical degrees of freedom.  
The results are interesting.  \\ 
In the fermionic sector, just three generations of quark and lepton states 
and some exotic states survive as shown in the following table.   

%
%
\begin{table}[htb]
\begin{center}
\begin{tabular}{|c|c|c|} \hline
\hspace{0mm} $SU(3)$ \hspace{0mm} & \hspace{0mm} $Q_e$ \hspace{0mm} & \hspace{0mm} 
$SU(2) \otimes U(1)$ \hspace{0mm} \\ \hline 
{{} 1} & 
$
\begin{array}{c}
0 \\[1mm]
-1 \\[1mm]
-2
\end{array} 
$ 
& \raise2ex\hbox{$\left(
\begin{array}{c}
{ \nu_e }    \\[1mm]
{{} e }  \\
\end{array} 
\right)$} 
\raise2ex\hbox{$\left(
\begin{array}{c}
{{} \nu_\mu} \\[1mm]
{{} \mu }
\end{array} 
\right)$} 
\raise2ex\hbox{$\left(
\begin{array}{c}
{{} \nu_\tau}   \\[1mm]
{{} \tau} 
\end{array} 
\right)$} 
\lower3ex\hbox{$
\begin{array}{c}
\left(
E
\right)
\end{array} 
$}
\\ \hline
{{} 3} & 
$
\begin{array}{c}
{5/3} \\[2mm]
{2/3} \\[2mm]
{-1/3} \\[2mm]
{-4/3}
\end{array} 
$ 
& \lower1ex\hbox{$\left(
\begin{array}{c}
{{} u }   \\[2mm]
{{} d }   \\
\end{array} 
\right)$} 
\lower1ex\hbox{$\left(
\begin{array}{c}
{{} c}  \\[2mm]
{{} s }  
\end{array} 
\right)$} 
\lower1ex\hbox{$\left(
\begin{array}{c}
{{} t } \\[2mm]
{{} b }   
\end{array} 
\right)$} 
\lower5ex\hbox{$\left(
\begin{array}{c}
h \\[1mm]
o
\end{array} 
\right)$}
\raise3ex\hbox{$\left(
\begin{array}{c}
a \\[2mm]
f
\end{array} 
\right)$}
\raise3ex\hbox{$\left(
\begin{array}{c}
g \\[1mm]
m
\end{array} 
\right)$}
\lower3ex\hbox{$\left(
\begin{array}{c}
r \\[1mm]
i \\[1mm]
n

\end{array} 
\right)$}
\\ \hline
{{} 6} & 
$
\begin{array}{c}
{4/3} \\[2mm]
{1/3} \\[2mm]
-{2/3}
\end{array} 
$ 
& $\left(
\begin{array}{c}
P \\[2mm]
Q \\[2mm]
R
\end{array} 
\right)$
$\left(
\begin{array}{c}
X \\[2mm]
Y \\[2mm]
Z
\end{array} 
\right)$
\\ \hline
{{} 8} & 
$
\begin{array}{c}
0 \\[2mm]
-1
\end{array} 
$ 
& $\left(
\begin{array}{c}
N_1 \\[2mm]
E_1
\end{array} 
\right)$
$\left(
\begin{array}{c}
N_2 \\[2mm]
E_2
\end{array} 
\right)$
\\ \hline
\end{tabular}
\end{center}
\end{table}
%
%
%
\noindent
The list of spin ${1 \over 2}$  survivours  after superHiggs-like mechanism are shown tentatively  as Dirac particles in the table.   \\
\indent
In the bosonic sector, gauge fields of SM in vector states and one Higgs scalar field of SM survive.
Besides those observed states,  one color-singlet neutral massive vector state $S$ and one color-singlet double-charge massive spin ${1 \over 2}$ state $E^{\pm 2}$are predicted,  which can be tested in the high energy (cosmic ray ) experiment. 
The simple extension of the model \cite{KS1} to larger $N>10$ produces  
massless charged high spin states,  ugly generation structures, etc. 
and is excluded.
%

%
We will show in the next section that no-go theorem for constructing non-trivial $SO(N > 8) $susy theory including gravity can be  circumvented by adopting the {\it nonliner (NL)} representation of SUSY \cite{WB}, 
i.e. by introducing {\it the degeneracy of space-time} through NLSUSY degrees of freedom. 
\section{ Nonlinear-Supersymmetric General Relativity (NLSUSYGR) }
For simplicity we discuss $N=1$ without the loss of the generality. 
The extension to $N>1$ is straightforward. 
The fundamental action 
{\it nonlinear supersymmetric general relativity theory (NLSUSYGR)} 
has been constructed by extending the geometric arguments 
of Einstein general relativity (EGR) on Riemann space-time to new space-time inspired by NLSUSY\cite{KS3,KS4}.  
The tangent space of new space-time is specified  not only by the Minkowski coodinate $x_a$ for $SO(1,3)$ 
but also by the Grassmann coordinate $\psi_\alpha$ for $SL(2,C)$ related to NLSUSY \cite{KS3,KS2}. 
They are coordinates of the coset space  ${superGL(4,R) \over GL(4,R)}$ 
and  can be interpreted as NG fermions associated with the spontaneous breaking of ${\it super}GL(4,R)$ down to $GL(4,R)$. 
(The  noncompact isomorphic groups $SO(1,3)$ and $SL(2,C)$ for tangent space-time symmetry on curved space-time 
can be regarded as the generalization of the compact isomorphic groups $SU(2)$ and $SO(3)$ for the gauge symmetry 
of 't Hooft-Polyakov monopole on flat space-time.)  
The NLSUSYGR action \cite{KS2,KS3}  is given by 
\begin{eqnarray}
& & L_{NLSUSYGR}(w) =- {c^4 \over 16{\pi}G} \vert w \vert \{\Omega(w) + \Lambda \}, 
\label{SGM}
\\[2mm]
& & \hspace*{7mm} 
\vert w \vert = \det w^a{}_\mu = \det \{e^a{}_\mu + t^a{}_\mu(\psi)\}, 
\nonumber \\[.5mm]
& & \hspace*{7mm} 
t^a{}_\mu(\psi) = {\kappa^2 \over 2i}(\bar\psi \gamma^a \partial_\mu \psi 
- \partial_\mu \bar\psi \gamma^a \psi), 
\label{Lw}
\end{eqnarray}
where $G$ is the Newton gravitational constant, $\Lambda$ is a ({\it small}) cosmological term and 
$\kappa$ is an arbitrary constant of NLSUSY with the dimemsion (mass)${^{-2}}$.   
$w^a{}_\mu(x)$ $= e^a{}_\mu + t^a{}_\mu(\psi)$ and 
$w^{\mu}{_a}$ = $e^{\mu}{_a}
- t{^{\mu}}_a + t{^{\mu}}_{\rho} t{^{\rho}}_a - t{^{\mu}}_{\sigma}t{^{\sigma}}_{\rho} t{^{\rho}}_a  
+ t{^{\mu}}_{\kappa}t{^{\kappa}}_{\sigma}t{^{\sigma}}_{\rho}t{^{\rho}}_a$  \
which terminate at $O(t^{4})$ for $N=1$ are the invertible {\it unified vierbeins} of new space-time.
{} $e^a{}_\mu$ is the ordinary vierbein of EGR for the  $SO(1,3)$ and    
$t^a{}_\mu(\psi)$ is the mimic vierbein analogue (actually the stress-energy-momentum tensor) of NG fermion $\psi(x){}$ for the $SL(2,C)$. 
(We call $\psi(x){}$  $superon$ as the hypothetical fundamental  spin $1 \over 2$  particle 
quantized canonically in compatible with  the sP algebra\cite{KS4}.) 
$\Omega(w)$ is the the unified Ricci scalar curvature of new space-time computed in terms of the {\it unified vierbein}  $w^a{}_\mu(x)$.  
Interestingly Grassmann degrees of freedom induce the imaginary part of 
the unified vierbein $w^a{}_\mu(x)$,  which represents straightforwardly the fermionic matter contribution.
Note that $e^a{}_\mu$ and $t^a{}_\mu(\psi)$ contribute equally to the curvature of spac-time,  which may be regarded as the Mach's principle in ultimate space-time. 
(The second index of mimic vierbein $t$, e.g. $\mu$ of $t^a{}_\mu$,  means the derivative $\partial_{\mu}$.)
$s_{\mu \nu} \equiv w^a{}_\mu \eta_{ab} w^b{}_\nu$ and 
$s^{\mu \nu}(x) \equiv w^\mu{}_a(x)w^{\nu a}(x)$ 
are {\it unified metric tensors} of new spacetime. 

NLSUSY GR action (\ref{SGM}) possesses promissing large symmetries 
isomorphic to $SO(N)$ ($SO(10)$) SP group \cite{ST3,ST4}, 
namely, $L_{NLSYSYGR}(w)$ is invariant under 
\begin{eqnarray}
 [{\rm new \ NLSUSY}] \otimes [{\rm local \ GL(4,R)}] \otimes [{\rm local \ Lorentz}] 
\end{eqnarray}
for space-time symmetries 
and 
\begin{equation}
[{\rm global} SO(N)] \otimes [{\rm local} U(1)^N] 
\end{equation}
for internal symmetries in case of $N$ superons $\psi^{i}, i=1,2,\cdots,N$.  \\
For example,  $L_{NLSUSYGR}(w)$ (\ref{Lw}) is invariant under the following NLSUSY transformations:
\begin{equation}
\delta^{NL} \psi ={1 \over \kappa^{2}} \zeta + 
i \kappa^{2} (\bar{\zeta}{\gamma}^{\rho}\psi) \partial_{\rho}\psi,
\quad
\delta^{NL} {e^{a}}_{\mu} = i \kappa^{2} (\bar{\zeta}{\gamma}^{\rho}\psi)\partial_{[\rho} {e^{a}}_{\mu]},
\label{newsusy}
\end{equation} 
where $\zeta$ is a constant spinor parameter and  $\partial_{[\rho} {e^{a}}_{\mu]} = 
\partial_{\rho}{e^{a}}_{\mu}-\partial_{\mu}{e^{a}}_{\rho}$, 
which induce the following GL(4,R) transformations on the unified vierbein  $w{^a}_{\mu}$ 
\begin{equation}
\delta_{\zeta} {w^{a}}_{\mu} = \xi^{\nu} \partial_{\nu}{w^{a}}_{\mu} + \partial_{\mu} \xi^{\nu} {w^{a}}_{\nu}, 
\quad
\delta_{\zeta} s_{\mu\nu} = \xi^{\kappa} \partial_{\kappa}s_{\mu\nu} +  
\partial_{\mu} \xi^{\kappa} s_{\kappa\nu} 
+ \partial_{\nu} \xi^{\kappa} s_{\mu\kappa}, 
\label{newgl4r}
\end{equation} 
where  $\xi^{\rho}=i \kappa^{2} (\bar{\zeta}{\gamma}^{\rho}\psi)$, \\ 
The commutators of two new NLSUSY transformations (\ref{newsusy})  on $\psi$ and  ${e^{a}}_{\mu}$ are $GL(4,R)$; 
\begin{equation}
[\delta_{\zeta_1}, \delta_{\zeta_2}] \psi
= \Xi^{\mu} \partial_{\mu} \psi,
\quad
[\delta_{\zeta_1}, \delta_{\zeta_2}] e{^a}_{\mu}
= \Xi^{\rho} \partial_{\rho} e{^a}_{\mu}
+ e{^a}_{\rho} \partial_{\mu} \Xi^{\rho},
\label{com1/2-e}
\end{equation}
where 
$\Xi^{\mu} = 2i\kappa (\bar{\zeta}_2 \gamma^{\mu} \zeta_1)
      - \xi_1^{\rho} \xi_2^{\sigma} e{_a}^{\mu}
      (\partial_{[\rho} e{^a}_{\sigma]})$.
The algebra closes. 
The ordinary local GL(4,R) invariance is trivial by the construction.  \\  
Also NLSUSYGR is invariant  under the following local Lorentz transformation:        \\      
on $w{^a}_{\mu}$ \\
\begin{equation}
\delta_L w{^a}_{\mu}
= \epsilon{^a}_b w{^b}_{\mu}
\label{Lrw}
\end{equation}
or equivalently on  $\psi$ and $e{^a}_{\mu}$
\begin{equation}
\delta_L \psi = - {i \over 2} \epsilon_{ab}
      \sigma^{ab} \psi,     \quad
\delta_L {e^{a}}_{\mu} = \epsilon{^a}_b e{^b}_{\mu}
      + {\kappa^{4} \over 4} \varepsilon^{abcd}
      \bar{\psi}\gamma_5 \gamma_d \psi
      (\partial_{\mu} \epsilon_{bc}),
\label{newlorentz}
\end{equation}
with the local  parameter $\epsilon_{ab} = (1/2) \epsilon_{[ab]}(x)$.    
The local Lorentz transformation forms a closed algebra, for example, on $e{^a}_{\mu}$ 
\begin{equation}
[\delta_{L_{1}}, \delta_{L_{2}}] e{^a}_{\mu}
= \beta{^a}_b e{^b}_{\mu}
+ {\kappa^{4} \over 4} \varepsilon^{abcd} \bar{\psi}
\gamma_5 \gamma_d \psi
(\partial_{\mu} \beta_{bc}),
\label{comLr1/2}
\end{equation}
where $\beta_{ab}=-\beta_{ba}$ is defined by
$\beta_{ab} = \epsilon_{2ac}\epsilon{_1}{^c}_{b} -  \epsilon_{2bc}\epsilon{_1}{^c}_{a}$.  \par
Note that the no-go theorem is overcome (circumvented) in a sense that 
the nontivial $N(N > 8)$-extended SUSY theory with the gravitational interaction  has been constructed in the global NLSUSY invariant way. \par
\section{Big Collapse of space-time}
New {\it (empty)} {} space-time  described  
by NLSUSYGR action $L_{NLSUSYGR}(w)$\cite{KS2,KS3} of the  EH-type 
equipping cosmological constant
is {\it unstable} due to NLSUSY structure of tangent space-time and 
would collapse (called {\it Big Collapse} \cite{ST4}) spontaneously to 
ordinary Riemann space-time with the cosmological constant and fermionic matter {\it superon}  ( called {\it superon-graviton model (SGM)} ). 
Note that Big Collapse induces inststantaneously the rapid expansion of three dimensional space-time due to the Pauli exclusion principle of NG fermion {\it superon}.   
SGM action is the following; 
\begin{eqnarray}
& & L_{SGM}(e, \psi) = -{c^4\Lambda \over 16{\pi}G} e \vert w \vert - {c^4 \over 16{\pi}G} e \vert w \vert R(e) 
+ {c^4 \over 16{\pi}G} e \vert w \vert \big[ \ 2 t^{(\mu\nu)} R_{\mu\nu}(e)    
\nonumber \\[.5mm]
& & + {1 \over 2} \{ g^{\mu\nu}\partial^{\rho}\partial_{\rho}t_{(\mu\nu)}
- t_{(\mu\nu)}\partial^{\rho}\partial_{\rho}g^{\mu\nu}       
 + g^{\mu\nu}\partial^{\rho}t_{(\mu\sigma)}\partial^{\sigma}g_{\rho\nu}
- 2g^{\mu\nu}\partial^{\rho}t_{(\mu\nu)}\partial^{\sigma}g_{\rho\sigma}
+\dots 
 \}     
\nonumber \\[.5mm]
& & + ({t^{\mu}}_{\rho}t^{\rho\nu}+{t^{\nu}}_{\rho}t^{\rho\mu}+t^{\mu\rho}{t^{\nu}}_{\rho})R_{\beta\mu}(e)    
- \{ 2 t^{(\mu\rho)}{t^{(\nu}}_{\rho)}R_{\mu\nu} + t^{(\mu \rho)} t^{(\nu \sigma)} R_{\mu \nu \rho \sigma}(e)  
\nonumber \\[.5mm]
& & + {1 \over 2}t^{(\mu\nu)}( g^{\rho\sigma}\partial_{\mu}\partial_{\nu}t_{(\rho\sigma)} 
- g^{\rho}{}_{\sigma}\partial^{\rho}\partial_{\mu}t_{(\sigma\nu)} + \dots )  \}  
 +\{ O(t^{3})\} + \cdots  \big],
\label{L-exp}
\end{eqnarray}
where $ \vert w \vert = det{w^{a}}_{b}=det(\delta^{a}{_b}+ t^{a}{_b}) $ , $e=det{e^{a}}_{\mu}$, $t^{(\mu\nu)}=t^{\mu\nu}+t^{\nu\mu}$, $t_{(\mu\nu)}=t_{\mu\nu}+t_{\nu\mu}$ and 
$(\psi)^{5} \equiv 0$(for $N=1$). $ \vert w \vert$  is the flat space NLSUSY action of VA\cite{VA} containing up to $O(t^{4})$ and $R(e)$, $R_{\mu\nu}(e)$, and $R_{\mu\nu\rho\sigma}(e)$ are the familiar curvature tensors of Riemann space of GR.   
Remarkably the first term should reduces to NLSUSY action \cite{VA} in Riemann-flat $e{_a}^{\mu}(x) \rightarrow \delta{_a}^{\mu}$ space-time, 
i.e. the arbitrary constant ${\kappa}$ of NLSUSY is fixed to  
\begin{equation}
{\kappa}^{-2} = {c^4 \over 8{\pi}G}{\Lambda}.
\label{kappa}
\end{equation}
$L_{SGM}(e, \psi)$ (\ref{L-exp}) can be recasted formally as the following famlliar form  
\begin{equation}
L_{SGM}(e,\psi)=-{c^{4} \over 16{\pi}G}\vert e \vert \{ R(e) + \Lambda + \tilde T(e, \psi) \},
\label{SGMR}
\end{equation}
where $R(e)$ is the Ricci scalar curvature of ordinary EH action and 
$\tilde T(e,\psi)$ represents the kinetic term and the gravitational interaction of superons.  \par
We have shown qualitatively that NLSUSYGR/SGM may describe a new paradigm for the SUSY unification of space-time and matter.  
Considering  that the graviton is the universal attractive force  
constituting all possible nontrivial composites(combinations) of superons, 
which is equivalent to the all possible  products of supercharges used in constructing the massless helicity representattion of sP group,  
The vacuum configuration of $L_{NLSUSYGR}(w) = L_{SGM}(e, \psi) $ may be achieved by producing gravitational composites of superons as the eigen states( ${\it  LSUSY supermultiplet}$)  of  sP space-time symmetry. 
That is, there may be a possibility that all (observed) low energy particles may be gravitational eigenstates of $SO(N)$ sP expressed uniquely 
as the SUSY composites of $N$ superons.  
We study explicitly these possibilities in the next section.
\section{Linearizing NLSUSY and vacuum of SGM}
\subsection{Linearization of NLSUSY}
The relation between the global LSUSY representation and the global NLSUSY one 
in flat space-time is studied in detail \cite {IK,R,UZ}. 
They have shown for $N=1$ SUSY in flat space-time that NSUISY can be linearized, i.e., the LSUSY transformation of a specific supermultiplet can be reproduced in terms of NLSUSY transformation of the NG field  and the (equivalent) relation of the two actions  is shown.     
We anticipate that  $L_{\rm SGM}(e,\psi)$ is linearized as well. 
Unfortunately due to the high nonlinearity of SGM action, 
the linearization of SGM actionn and extracting the (low energy)  physical meaning of SGM directly on curved Riemann space-time is yet to be done.   \\     
%
However, 
considering that the SGM action  in 
Riemann-flat $(e^a{}_\mu \rightarrow \delta^a_\mu)$ space-time reduces essentially to the $N$-extended NLSUSY action 
with ${\kappa^{2} = ({c^{4}\Lambda \over 8{\pi}G}})^{-1}$,   
it is interesting from the viewpoint of the low energy physics on the tangent flat space-time to linearize the  $N$-extended NLSUSY model and find  the equivqlent(related) $N$-extended LSUSY theory.  
We show explicitly in two dimensional space-time ($d = 2$) \cite{ST2,ST5} for simplicity that $N = 2$ 
LSUSYQED is equivalent(related) to $N = 2$ NLSUSY model. \par
Firstly we perform the  heuristic approach of the linearization based on the commutator 
of SUSY algebra.     
The heuristic approach  is suggested by the following observations and 
gives the intuitive understandings of the linearization 
which is  formulated on the  various Lorentz tensors 
and  ${ \vert w  \vert}$\cite{UZ,T1}                 \par 
The product of {Lorentz tensors composed of  $\psi^{i}$ 
multiplied by ${ \vert w \vert}$\cite{T1}}:   \\
\ba
b^i{}_A{}^{jk}{}_B{}^{l \cdots m}{}_C{}^n \left( (\psi^i)^{2(n-1)} \vert w \vert \right) 
\A = \A \kappa^{2n-3} \bar\psi^i \gamma_A \psi^j \bar\psi^k \gamma_B \psi^l \cdots \bar\psi^m \gamma_C \psi^n \vert w \vert, 
\label{bosonic} 
\\[1mm]
f^{ij}{}_A{}^{kl}{}_B{}^{m \cdots n}{}_C{}^p \left( (\psi^i)^{2n-1} \vert w \vert \right) 
\A = \A \kappa^{2(n-1)} \psi^i \bar\psi^j \gamma_A \psi^k \bar\psi^l \gamma_B \psi^m \cdots \bar\psi^n \gamma_C \psi^p \vert w \vert, 
\label{fermionic}
\ea
play   basic roles.    \\%

The variations under the NLSUSY transformations become
%
\ba
\delta_\zeta { { b^i{}_A{}^{jk}{}_B{}^{l \cdots m}{}_C{}^n}} 
\A = \A 
\kappa^{2(n-1)} \left[ \left\{ \left( \bar\zeta^i \gamma_A \psi^j + \bar\psi^i \gamma_A \zeta^j \right) 
\bar\psi^k \gamma_B \psi^l \cdots \bar\psi^m \gamma_C \psi^n + \cdots \right\} \vert w \vert \right. 
\nonu
\A \A 
\left. + \kappa \partial_a \left( \xi^a \bar\psi^i \gamma_A \psi^j \bar\psi^k \gamma_B \psi^l 
\cdots \bar\psi^m \gamma_C \psi^n \vert w \vert \right) \right], 
\label{variation1}
\\[1mm]
\delta_\zeta f^{ij}{}_A{}^{kl}{}_B{}^{ml \cdots n}{}_C{}^p
\A = \A 
\kappa^{2n-1} \left[ \left\{ \zeta^i \bar\psi^j \gamma_A \psi^k \bar\psi^l \gamma_B \psi^m \cdots \bar\psi^n \gamma_C \psi^p 
\right. \right. 
\nonu
\A \A 
\left. + \psi^i \left( \bar\zeta^j \gamma_A \psi^k + \bar\psi^j \gamma_A \zeta^k \right) 
\bar\psi^l \gamma_B \psi^m \cdots \bar\psi^n \gamma_C \psi^p + \cdots \right\} \vert w \vert 
\nonu
\A \A 
\left. + \kappa \partial_a \left( \xi^a \psi^i \bar\psi^j \gamma_A \psi^k \bar\psi^l \gamma_B \psi^m 
\cdots \bar\psi^n \gamma_C \psi^p \vert w \vert \right) \right], 
\label{variation2}
\ea
where 
$\xi^{a}=i\kappa\bar\zeta^i \gamma^{a} \psi^i.$    \\[3mm]
These show that 
{Lorentz tensors of $\psi^{i}$ miltiplied by ${\vert w \vert}$} \\
 {give} { {{finite}} representation of NLSUSY algebra}, 
 because of  $({\psi^{i}}_\alpha)^{n} \equiv 0,  n >4N. $  \\  

\noindent
They satisfy the commutator
\begin{equation}
[ \delta_Q(\zeta_1), \delta_Q(\zeta_2)] 
= \delta_P(v), 
\label{commutator}
\end{equation}
where $\delta_P(v)$ is a translation with a parameter 
$v^a = 2i ( \bar\zeta_{1L}^{i} \gamma^a \zeta_{2L}^{i} 
- \bar\zeta_{1R}^{i} \gamma^a \zeta_{2R}^{i} ).$   \\
These results show that the commutator-based linearization closes on the all possible Lorents tensors composed of $\psi^{i}$ and gives a finite dimensional representation of sP albebra.    \\
In the heuristic commutator based linearization 
we consider  two  steps:        \\ 
(i) {\it SUSY compositenes based on LSUSY algebra}: \ \
Find composite LSUSY  supermultiplet,  i.e.  every component  field including  the auxiliary field  of LSUSY supermultiplet  should be expressed  as the Lorentz tensors composed of the products of the NLSUSY NG fermion $\psi$  and simultaneously  the  familiar LSUSY transformation on the supermultiplet should be reproduced(satisfied) in terms of  the composite supermultiplet under the NLSUSY transformations of the constituent NG  fermion $\psi$,  \\ 
(ii) {\it NL/L SUSY relation(equivalence)}: \ Show   LSUSY action $L_{LSUSY}$ reduces to NLSUSY  action  $L_{NLSUSY}$ when  {\it SUSY compositeness} is substituted into  LSUSY field of  $L_{LSUSY}$.  \par 
Now we consider  explicitly the vacuum structure of $N = 2$ LSUSY QED in the SGM scenario in $d = 2$ \cite{ST5}, which enables to study the vacuum structure of SGM scenario. 
(Note that the minimal realistic SUSY QED in SGM composite scenario is given by $N = 2$ SUSY \cite{STT2}.) 
$N = 2$
NLSUSY action for two superons (NG fermions) $\psi^i\ (i=1,2)$ in $d = 2$ is written as follows, 
\begin{eqnarray}
& & L_{N=2{\rm NLSUSY}}
\nonumber \\[.5mm]
& & = -{1 \over {2 \kappa^2}} \vert w \vert
\nonumber \\[.5mm]
& & = - {1 \over {2 \kappa^2}} 
\left\{ 1 + t^a{}_a + {1 \over 2!}(t^a{}_a t^b{}_b - t^a{}_b t^b{}_a) 
\right\} 
\nonumber \\[.5mm]
& & = - {1 \over {2 \kappa^2}} 
\left\{ 1 - i \kappa^2 \bar\psi^i \!\!\not\!\partial \psi^i \right. 
\nonumber \\[.5mm]
& & 
\left. - {1 \over 2} \kappa^4 
( \bar\psi^i \!\!\not\!\partial \psi^i \bar\psi^j \!\!\not\!\partial \psi^j 
- \bar\psi^i \gamma^a \partial_b \psi^i \bar\psi^j \gamma^b \partial_a \psi^j ) 
\right\} 
\label{VAaction2}
\end{eqnarray}
where $\kappa$ is a constant whose dimension is $({\rm mass})^{-1}$ and 
$\vert w \vert = \det(w^a{}_b) = \det(\delta^a_b + t^a{}_b)$, 
$t^a{}_b = - i \kappa^2 \bar\psi^i \gamma^a \partial_b \psi^i$. 
While, the helicity states of  $N = 2$ LSUSY QED are 
the vector supermultiplet containing $U(1)$ gauge field 
\[\left(\begin{array}{c}
      +{1} \\
\begin{array}{cc}
 +{1 \over 2}, \  +{1 \over 2} 
\end{array} \\
0
\end{array}  \right) + [{\rm CPT\ conjugate}], \]
and the scalar supermultiplet for matter fields  
\[\left(\begin{array}{c}
      +{1 \over 2} \\
\begin{array}{cc}
 0 ,\  0 
\end{array} \\
-{1 \over 2}
\end{array}  \right) + [{\rm CPT\ conjugate}]. \]
The most general $N = 2$ LSUSY QED action in the Wess-Zumino gauge for the massless case in $d = 2$, 
is written as follows\cite{ST5}, 
\begin{eqnarray}
& & L_{N=2{\rm SUSYQED}} 
\nonumber \\[.5mm]
& & = - {1 \over 4} (F_{ab})^2 
+ {i \over 2} \bar\lambda^i \!\!\not\!\partial \lambda^i 
+ {1 \over 2} (\partial_a A)^2 
+ {1 \over 2} (\partial_a \phi)^2 
+ {1 \over 2} D^2 
\nonumber \\[.5mm]
& & 
- {1 \over \kappa} \xi D 
+ {i \over 2} \bar\chi \!\!\not\!\partial \chi 
+ {1 \over 2} (\partial_a B^i)^2 
+ {i \over 2} \bar\nu \!\!\not\!\partial \nu 
+ {1 \over 2} (F^i)^2 
\nonumber \\[.5mm]
& & 
+ f ( A \bar\lambda^i \lambda^i + \epsilon^{ij} \phi \bar\lambda^i \gamma_5 \lambda^j 
- A^2 D + \phi^2 D + \epsilon^{ab} A \phi F_{ab} ) 
\nonumber \\[.5mm]
& & 
+ e \left\{ i v_a \bar\chi \gamma^a \nu 
- \epsilon^{ij} v^a B^i \partial_a B^j 
+ \bar\lambda^i \chi B^i 
+ \epsilon^{ij} \bar\lambda^i \nu B^j 
\right. 
\nonumber \\[.5mm]
& & 
\left. 
- {1 \over 2} D (B^i)^2 
+ {1 \over 2} (\bar\chi \chi + \bar\nu \nu) A 
- \bar\chi \gamma_5 \nu \phi \right\}
\nonumber \\[.5mm]
& & 
+ {1 \over 2} e^2 (v_a{}^2 - A^2 - \phi^2) (B^i)^2, 
\label{L2action}
\end{eqnarray}
where  $(v^a, \lambda^i, A, \phi, D)$ ($F_{ab} = \partial_a v_b - \partial_b v_a$) 
is the  {\it minimal} off-shel vector supermultiplet containing $v^a$ for a $U(1)$ vector field, 
$\lambda^i$ for doublet (Majorana) fermions,  
$A$ for a scalar field in addition to $\phi$ for another scalar field 
and $D$ for an auxiliary scalar field,  
while ($\chi$, $B^i$, $\nu$, $F^i$) is the {\it minimal} off-shell scalar supermultiplet containing 
$(\chi, \nu)$ for two (Majorana) fermions, 
$B^i$ for doublet scalar fields and $F^i$ for auxiliary scalar fields. 
The linear term of $F$ is forbidden by the gauge invariance\cite{ST5}. 
Also $\xi$ is an arbitrary demensionless parameter giving a magnitude of SUSY breaking mass, 
and $f$ and $e$ are Yukawa and gauge coupling constants with the dimension (mass)$^1$, respectively. \par
$N = 2$ LSUSY QED action (\ref{L2action}) is invariant under the following LSUSY transformations parametrized by $\zeta^i$, 
\begin{eqnarray}
& &
\delta_\zeta v^a = - i \epsilon^{ij} \bar\zeta^i \gamma^a \lambda^j,  \
\delta_\zeta \lambda^i 
= (D - i \!\!\not\!\partial A) \zeta^i 
+ {1 \over 2} \epsilon^{ab} \epsilon^{ij} F_{ab} \gamma_5 \zeta^j 
- i \epsilon^{ij} \gamma_5 \!\!\not\!\partial \phi \zeta^j,   \\
\nonumber
& &
\delta_\zeta A = \bar\zeta^i \lambda^i,  \
\delta_\zeta \phi = - \epsilon^{ij} \bar\zeta^i \gamma_5 \lambda^j,   \
\delta_\zeta D = - i \bar\zeta^i \!\!\not\!\partial \lambda^i. 
\label{VLSUSY}
\end{eqnarray}
for the vector multiplet and 
\begin{eqnarray}
& &
\delta_\zeta \chi 
 = (F^i - i \!\!\not\!\partial B^i) \zeta^i - e \epsilon^{ij} V^i B^j,  \
\delta_\zeta \nu 
=  \epsilon^{ij} (F^i + i \!\!\not\!\partial B^i) \zeta^j + e V^i B^i,  \\
\nonumber 
& &
\delta_\zeta B^i 
=  \bar\zeta^i \chi - \epsilon^{ij} \bar\zeta^j \nu,  \\
& &
\delta_\zeta F^i 
 = - i \bar\zeta^i \!\!\not\!\partial \chi 
- i \epsilon^{ij} \bar\zeta^j \!\!\not\!\partial \nu 
\nonumber 
- e \{ \epsilon^{ij} \bar V^j \chi - \bar V^i \nu 
+ (\bar\zeta^i \lambda^j + \bar\zeta^j \lambda^i) B^j 
- \bar\zeta^j \lambda^j B^i \} 
\label{SLSUSYg}
\end{eqnarray}
%
with $V^i = i v_a \gamma^a \zeta^i - \epsilon^{ij} A \zeta^j - \phi \gamma_5 \zeta^i$ for the scalar multiplet.  \par 
For extracting the low energy  physical contents of $N = 2$ SGM (NLSUSY GR) 
we consider SGM in asymptotic  Riemann-flat space-time, where $N = 2$ SGM reduces to essentially $N = 2$ NLSUSY action. 
We will show the relation(equivalence) of $N = 2$ NLSUSY action and 
$N = 2$ SUSYQED action(called {\it NL/L SUSY relation}.) :  
\begin{equation}
L_{N=2{\rm SGM}} \longrightarrow{(e^a{}_\mu \rightarrow \delta^a_\mu)}{\longrightarrow} 
L_{N=2{\rm NLSUSY}} + [{\rm suface\ terms}]
= f_{\xi} L_{N=2{\rm SUSYQED}}, 
\end{equation}
where $f_{\xi}$ is the function of  vacuum values of auxiliary fields.
NL/L SUSY relation indicates the equivalence(relation) of two theories  irrespective of the renormalizability.
NL/L SUSY relation is shown explicitly by substituting 
the following SUSY compositeness condition\cite{ST5} into the LSUSY QED theory. \par 
Now we find the SUSY compositeness of LSUSY gauge multiplet 
as follows.    \par 
Starting  tentatively  from  natural and  simple SUSY compositeness for 
$v^{a}, A, \lambda$ as 
$v^a =  \xi \kappa \epsilon^{ij} 
\bar\psi^i \gamma^a \psi^j \vert w \vert,,  \
A = {1 \over 2} \xi \kappa \bar\psi^i \psi^i \vert w \vert,  \
\lambda^i = \xi \psi^i \vert w \vert $ and  then operating NLSUSY transformation on the constituents $\psi$, the results contains various 
Lorentz tensors composed of $\psi.$. Comparing these results with the familiar LSUSY transformation of the gauge supermultiplet including the auxiliary field, we get the SUSY cmpositeness.   \par
The SUSY compositeness  condition for the minimal vector supermultiplet \\  $(v^a, \lambda^i, A, \phi, D)$ in the Wess-Zumino gauge\cite{ST1} are 
\begin{eqnarray}
& & 
v^a = - {i \over 2} \xi \kappa \epsilon^{ij} 
\bar\psi^i \gamma^a \psi^j \vert w \vert, 
\nonumber \\[.5mm]
& & 
\lambda^i = \xi \left[ \psi^i \vert w \vert 
- {i \over 2} \kappa^2 \partial_a 
\{ \gamma^a \psi^i \bar\psi^j \psi^j 
(1 - i \kappa^2 \bar\psi^k \!\!\not\!\partial \psi^k) \vert w \vert\} \right], 
\nonumber \\[.5mm]
& & 
A = {1 \over 2} \xi \kappa \bar\psi^i \psi^i \vert w \vert, 
\nonumber \\[.5mm]
& & 
\phi = - {1 \over 2} \xi \kappa \epsilon^{ij} \bar\psi^i \gamma_5 \psi^j 
\vert w \vert, 
\nonumber \\[.5mm]
& & 
D = {\xi \over \kappa} \vert w \vert 
- {1 \over 8} \xi \kappa^3 
\partial_a \partial^a ( \bar\psi^i \psi^i \bar\psi^j \psi^j\vert w \vert ).  
\label{SSUSYinv1}
\end{eqnarray}
While  for the minimal scalar supermultiplet $(\chi, B^i, \nu, F^i)$ 
the SUSY compositeness condition is   
\begin{eqnarray}
& & 
\chi = \xi^i \left[ \psi^i \vert w \vert 
+ {i \over 2} \kappa^2 \partial_a 
\{ \gamma^a \psi^i \bar\psi^j \psi^j 
(1 - i \kappa^2 \bar\psi^k \!\!\not\!\partial \psi^k)\vert w \vert \} \right], 
\nonumber  \\[.5mm]
& & 
B^i = - \kappa \left( {1 \over 2} \xi^i \bar\psi^j \psi^j 
- \xi^j \bar\psi^i \psi^j \right) \vert w \vert, 
\nonumber \\[.5mm]
& & 
\nu = \xi^i \epsilon^{ij} \left[ \psi^j \vert w \vert 
+ {i \over 2} \kappa^2 \partial_a 
\{ \gamma^a \psi^j \bar\psi^k \psi^k 
(1 - i \kappa^2 \bar\psi^l \!\!\not\!\partial \psi^l) \vert w \vert\} \right], 
\nonumber \\[.5mm]
& & 
F^i = {1 \over \kappa} \xi^i \left\{ \vert w \vert 
+ {1 \over 8} \kappa^3 
\partial_a \partial^a ( \bar\psi^j \psi^j \bar\psi^k \psi^k \vert w \vert) 
\right\} 
\nonumber \\[.5mm]
& & 
- i \kappa \xi^j \partial_a ( \bar\psi^i \gamma^a \psi^j \vert w \vert ) 
- {1 \over 4} e \kappa^2 \xi \xi^i \bar\psi^j \psi^j \bar\psi^k \psi^k\vert w \vert, 
\label{SSUSYinv}
\end{eqnarray}
where $\xi$ and $\xi^{i}$ are  factors of the vacuum expectation values of auxiliary fields $D$ and $F^i$.
SUSY compositeness  (\ref{SSUSYinv1}) and (\ref{SSUSYinv})  
satisfy  (i) and (ii).
Furthermore substituting these relations into the $N=2$ SUSYQED action (\ref{L2action})
we can show directly  NL/L SUSY  relation (19) , i.e.,   
$N=2$ SUSYQED action (\ref{L2action}) is equivalent(related) to $N=2$ NLSUSY action provided 
\begin{equation}
f_{\xi}(\xi,\xi^{i})=\xi^{2}-(\xi^{i})^{2}=1. 
\label{f}
\end{equation}
Note that for the SUSY compositeness (\ref{SSUSYinv}) of the scalar supermultiplet 
it is interesting that the four-fermion self-interaction term 
appearing only in the auxiliary fields $F^i$ 
is the origin of the familiar local $U(1)$ gauge symmetry of LSUSY theory. 
The introduction of  new auxiliary fields in the  supermultiplet improves(clarifies) these  situations. 
Especially the total derivative term in SUSY compositeness (\ref{SSUSYinv1}) and (\ref{SSUSYinv}) disappears  by introducing a new  auxiliary fields   composed of fermion self-interactions. 
Note that in the straightforward linearization the commutator algebra does not contain U(1) gauge transformation even for the vector U(1) gauge field\cite{STT2}. 
NL/L SUSY relation of SGM scenario for the larger supermultiplet predicts the magnitude of the bare gauge coupling constant as shown below.  
These situations are easily seen by using the superfield formulation of the linearizion of NLSUSY\cite{ST5a}.    \par
Now we discuss the linearization by the superfield formalism\cite{ST5}. 
We adopt for the $N = 2$ general vector supermultiplet of the general superfield  
\begin{eqnarray}
{\cal V}(x, \theta^i) &\!\!\! = &\!\!\! C(x) + \bar\theta^i \Lambda^i(x) 
+ {1 \over 2} \bar\theta^i \theta^j M^{ij}(x) 
- {1 \over 2} \bar\theta^i \theta^i M^{jj}(x) 
+ {1 \over 4} \epsilon^{ij} \bar\theta^i \gamma_5 \theta^j \phi(x) 
\nonumber \\[.5mm]
&\!\!\! &\!\!\! 
- {i \over 4} \epsilon^{ij} \bar\theta^i \gamma_a \theta^j v^a(x) 
- {1 \over 2} \bar\theta^i \theta^i \bar\theta^j \lambda^j(x) 
- {1 \over 8} \bar\theta^i \theta^i \bar\theta^j \theta^j D(x), 
\end{eqnarray}
and for the $N = 2$ scalar supermultiplet  
\begin{eqnarray}
\Phi^i(x, \theta^i) &\!\!\! = &\!\!\! B^i(x) + \bar\theta^i \chi(x) - \epsilon^{ij} \bar\theta^j \nu(x) 
- {1 \over 2} \bar\theta^j \theta^j F^i(x) + \bar\theta^i \theta^j F^j(x) 
- i \bar\theta^i \!\!\not\!\partial B^j(x) \theta^j 
\nonumber \\[.5mm]
&\!\!\! &\!\!\!
+ {i \over 2} \bar\theta^j \theta^j (\bar\theta^i \!\!\not\!\partial \chi(x) 
- \epsilon^{ik} \bar\theta^k \!\!\not\!\partial \nu(x)) 
+ {1 \over 8} \bar\theta^j \theta^j \bar\theta^k \theta^k \partial_a \partial^a B^i(x),  
\end{eqnarray}
\noindent
The N=2 SUSYQED action and its NL/L SUSY relation  by the superfield is expressed as: \\
\begin {equation}
S_{N = 2{\rm NLSUSY}},=S_{{\cal V}{\rm free}} + S_{{\cal V}f} +  S_{\rm gauge} 
= f_{\xi}(\xi,\xi^{i}, \xi_{c})S_{N = 2{\rm NLSUSY}}.
\label{N2relation}
\end{equation}
$S_{{\cal V}{\rm free}}$  is the  kinetic (free) action with the Fayet-Iliopoulos (FI) $D$ term 
for the $N = 2$ vector supermultiplet ${\cal V}$
\begin{eqnarray}
S_{{\cal V}{\rm free}} &\!\!\! = &\!\!\! \int d^2 x \left\{ 
\int d^2 \theta^i \ {1 \over 32} (\overline{D^i {\cal W}^{jk}} D^i {\cal W}^{jk} 
+ \overline{D^i {\cal W}_5^{jk}} D^i {\cal W}_5^{jk}) 
\ { + \int d^4 \theta^i \ {\xi \over {2 \kappa}} {\cal V}} 
\right\}_{\theta^i = 0} 
\end{eqnarray}
where 
\begin{equation}
{\cal W}^{ij} = \bar D^i D^j {\cal V}, \ \ \ {\cal W}_5^{ij} = \bar D^i \gamma_5 D^j {\cal V}. 
\end{equation}
\vspace{1mm}
\noindent
The FI $D$ term gives { the correct sign} of the NLSUSY action.  \par
%
{Yukawa interaction terms} for ${\cal V}$ is  \\ 
\begin{eqnarray}
S_{{\cal V}f} &\!\!\! = &\!\!\! {1 \over 8} \int d^2 x \ f \left[ 
\int d^2 \theta^i \ {\cal W}^{jk} 
({\cal W}^{jl} {\cal W}^{kl} + {\cal W}_5^{jl} {\cal W}_5^{kl}) 
\right. 
\nonumber \\[.5mm]
&\!\!\! &\!\!\! 
+ \int d \bar\theta^i d \theta^j \ 2 \{ {\cal W}^{ij} 
({\cal W}^{kl} {\cal W}^{kl} + {\cal W}_5^{kl} {\cal W}_5^{kl}) 
+ {\cal W}^{ik} ({\cal W}^{jl} {\cal W}^{kl} + {\cal W}_5^{jl} {\cal W}_5^{kl}) \} 
\bigg]_{\theta^i = 0} 
\end{eqnarray}
By means of {cancelations among four NG-fermion self-interaction terms}  
%
{General mass terms for $\tilde{\cal V}$ vanishes and $\tilde\Phi$}  as well. 
\par
%
%
The {\it most general} $U(1)$  gauge invariant action 
for ${\Phi^i}$  coupled with  ${\cal V}$ 
\begin{eqnarray}
S_{\rm gauge} &\!\!\! = &\!\!\! - {1 \over 16} \int d^2 x \int d^4 \theta^i 
e^{-4e{\cal V}} (\Phi^j)^2.
\end{eqnarray}       \par
%
%
Now we consider the following 
{ $\psi^i$}{-dependent superspace supertranslations}  \\ 
with ${ -{\kappa}{\psi(x)}}$, 
\begin{equation}
x'^a = x^a + i \kappa \bar\theta^i \gamma^a {{} \psi^i}, 
\ \ \ 
\theta'^i = \theta^i - \kappa {{} \psi^i}, 
\end{equation}
%
and denote the resulting superfields on $(x'^a, \theta'^i)$ and 
their $\theta$-expansions as 
\begin{equation}
{\cal V}(x'^a, \theta'^i) = \tilde{\cal V}(x^a, \theta^i; {{} \psi^i(x)}), 
\ \ \ 
\Phi(x'^a, \theta'^i) = \tilde\Phi(x^a, \theta^i; {{} \psi^i(x)}). 
\end{equation}
and consider the supertranslation of  $x',\theta'$, we obtain 
\begin{equation}
\delta\tilde{\cal V}(x^a, \theta^i; {{} \psi^i(x)}) = \xi_{\mu}\partial^{\mu}\tilde{\cal V}(x^a, \theta^i; {{} \psi^i(x)}), \
\delta\tilde{\Phi}(x^a, \theta^i; {{} \psi^i(x)}) = \xi_{\mu}\partial^{\mu}\tilde{\Phi}(x^a, \theta^i; {{} \psi^i(x)}), 
\end{equation}
where $\xi_{\mu}=-i{\bar{\varepsilon}}^{i}\gamma_{\mu}\psi^{i}$.
\noindent
Therefore, we obtain the following {generalized global  SUSY invariant constraints} 
for (the component fields in)  $\tilde{\cal V}$ and $\tilde\Phi$, 
\begin{equation}
\tilde\varphi_{\cal V}^I(x) = \xi^{I}_{\cal V}({\rm constant}) 
\ \ \
\tilde\varphi_\Phi^I(x) = \xi^{I}_{\Phi}({\rm constant}).  
\label{SUSYconst-SSF}
\end{equation}
Solving these equations  with respect to the initial components of supermultiplet  we obtain the SUSY compositeness conditions 
in terms of $\tilde\varphi_{\cal V}^I(x)$, $\tilde\varphi_{\Phi }^I(x)$ 
and $\psi^{i}(x)$.   \\
Now we impose the constraints from the physical viewpoint, i.e. the Lorentz invariance of the vacuum expectation value of auxiliary scalar field, the following simple SUSY invariant constraint is the minimal one:  
\begin{eqnarray}
\tilde C = \tilde\Lambda^i = \tilde M^{ij} = \tilde\phi = \tilde v^a = 
\tilde\lambda^i = 0, \ \
\tilde D ={\xi \over \kappa}, 
\tilde B^i = \tilde\chi = \tilde\nu = 0, \ \ \ \tilde F^i = {{{} \xi^i \over \kappa}}, 
\end{eqnarray}
which produces the previous simple SUSY compositeness  condition (\ref{SSUSYinv1}) and (\ref{SSUSYinv}) and NL/L SUSY relation(\ref{N2relation}) with (\ref{f}). 
By changing the integration variables 
{$(x^a, \theta^i)$ $\rightarrow$ $(x'^a, \theta'^i)$}, NL/L SUSY relation 
can be confirmed straightforwardly.  \\
More  general and physical  SUSY invariant constraint corresponding to vev of general scalar auxiliary fields: 
\begin{eqnarray}
{ {\tilde C ={\xi_{c}}}}, \ \tilde\Lambda^i = \tilde M^{ij} = \tilde\phi = \tilde v^a = 
\tilde\lambda^i= 0, \ \ {}
\tilde D = { \xi \over \kappa}, \ \
\tilde B^i = \tilde\chi = \tilde\nu = 0, \ \ \ \tilde F^i = {\xi^i \over \kappa}. 
\end{eqnarray}
gives SUSY compositeness condition for largear supermultiplet and produces for NL/L SUSY relation\cite{ST5a} 
\begin{equation}
f_{\xi}(\xi,\xi^{i}, \xi_{c})=\xi^2 - (\xi^i)^2e^{-4e\xi_{c}} = 1,  \  i.e.,  \
{e} = {{ \ln({\xi^i{}^2 \over {\xi^2 - 1}}) \over 4 \xi_c}}. 
\label{f-xi}
\end{equation}
Remarkably the bare gauge coupling constant  ${ e}$ is determined in terms of the the vev of the 
scalar fields.   
This mechanism seems natural and favorable from the viewpoint of theory of  everything. \par
\subsection{Vacuum structure of NLSUSY}
NL/L SUSY relation of N=2 SUSY means that 
the  physical configuration of the vacuum of NLSUSY theory 
is given by the the vacuum structure of $N = 2$ SUSY QED action (\ref{L2action}) \cite{STL}.
The vacuum of  $N = 2$ SUSY QED action is given by the minimum of the potential {{} {$V(A, \phi, B^i, D)$ of  $L_{N=2{\rm LSUSYQED}}$}}, 
\begin{equation}
V(A, \phi, B^i, D)  =  - {1 \over 2} D^2 + \left\{ {\xi \over \kappa} 
- f(A^2 - \phi^2) + {1 \over 2} e |B^i|^2 \right\} D 
+{e^{2} \over 2}(A^2 + \phi^2)|B^{i}|^{2}.   
\end{equation}
%
%
\noindent
{} Substituting the solution of the equation of motion for the auxiliary field $D$ we obtain
%
\begin{equation}
V(A, \phi, B^i) = {1 \over 2} f^2 \left\{ A^2 - \phi^2 - {e \over 2f} |B^i|^2 
- {\xi \over {f \kappa}} \right\}^2 +{1 \over 2} e^2 ( A^2 + \phi^2 )|B^i|^2 \ge 0. 
\end{equation}
%
%
The configuration of the fields corresponding to the vacua in $(A, \phi, B^i)$-space which are $SO(1,3)$ symmetric  classified according to the signatures of the parameters $e, f, \xi, \kappa$.  \par
We obtain two different types of vacua  $V=0$: \  
(I)$ {} \ \ {A=\phi=0}, \ \ |\tilde B^i|^2 =- k^2 $ 
 \ \ and  \ \ 
(II) ${} \ \ |\tilde B^i|^2 =0, \ \ A^2 - \phi^2  = k^2, 
 \ \left( \tilde B^i = \sqrt{e \over 2f} B^i, \ \ 
k^2 = {\xi \over {f \kappa}} \right)$. 
Parametrizing the potential configuration as
for case (I), \  
%
$A  = - (k + \rho) \cos\theta \cos\varphi \cosh\omega$,   \
$\phi  = (k + \rho) \sinh\omega$,  \   
$\tilde B^1  = (k + \rho) \sin\theta \cosh\omega$,    \   
$\tilde B^2  = (k + \rho) \cos\theta \sin\varphi \cosh\omega$.     
%
%
and \ 
for case (II):, \ 
%
$A  = (k + \rho)\sin\theta \cosh\omega$,   \      
$\phi = (k + \rho) \sinh\omega$        \  
$\tilde B^1  = (k + \rho) \cos\theta \cos\varphi \cosh\omega$,    \  
$\tilde B^2  = (k + \rho) \cos\theta \sin\varphi \cosh\omega$.     \
\noindent
Expanding  fields around the vacuum values of the parameter space
and substituting these expressions into $ L_{N=2{\rm SUSYQED}}$ $(A, \phi, B^i)$  we obtain the physical particle configuration around the true vacuum. 
For example,  we have for the type (II) vacuum
\begin{eqnarray}
& & L_{N=2{\rm SUSYQED}} 
\nonumber \\
&=& {1 \over 2} \{ (\partial_a \hat A)^2 - 4 f^2 k^2 \hat A^2 \} 
+ {1 \over 2} \{ |\partial_a \hat B^1|^2 + |\partial_a \hat B^2|^2 
- e^2 k^2 (|\hat B^1|^2 + |\hat B^2|^2) \} 
%
+ {1 \over 2} (\partial_a \hat \phi)^2 
\nonumber  \\
& & 
- {1 \over 4} (F_{ab})^2 
+ {1 \over 2} (i \bar\lambda^i \!\!\not\!\partial \lambda^i 
- 2 f k \bar\lambda^i \lambda^i) 
+ {1 \over 2} \{ i (\bar\chi \!\!\not\!\partial \chi + \bar\nu \!\!\not\!\partial \nu) 
- e k (\bar\chi \chi + \bar\nu \nu) \} 
+ \cdots. 
\end{eqnarray}
%
%
%
and the following mass spectra of particles(tilded fields below) in the true vacuum: \\ 
\begin{eqnarray}%
&&
m_{\hat A}^2 = m_{\lambda^i}^2 = 4 f^2 k^2 = {{4 \xi f} \over \kappa},  \ \
\nonumber  \\
& & 
m_{\hat B^1}^2 = m_{\hat B^2}^2 = m_\chi^2 = m_\nu^2 
= e^2 k^2 = {{\xi e^2} \over {\kappa f}},   \  \ \
\nonumber  \\
& & 
m_{v_{a}} = m_{\hat \phi} = 0.     \\   
\nonumber
\end{eqnarray}
with the mass hierarchy by the factor ${e \over f}$. 
The non-zero vacuum value of the scalar field breaks SUSY and gives 
mass to  the fermion.  
The resulting model describes: \\
one massive charged Dirac fermion ({$\psi_D{}^c \sim \chi + i \nu$}), 
one massive neutral Dirac fermion ({$\lambda_D{}^0 \sim \lambda^1 - i \lambda^2$}), 
one massless vector (a photon) ({{} $v_a$}), 
one charged scalar ({{} $\hat B^1 + i \hat B^2$}), 
one neutral complex scalar ({{} $\hat A+ i \hat \phi $}),  $\hat \phi$ 
is Goldstone mode in SO(1.3) plane, 
which are {{} composites of superons}. \par
Apparently the identification of the supersymmetric partners 
is not manifest in the resulting spectra.   
({without manifest superpartners}).
From these arguments we conclude that $N = 2$ SUSY QED is equivalent(related) to $N = 2$ NLSUSY action, i.e., to the matter sector of 
$N = 2$ SGM produced by Big Collapse of new space-time 
$N = 2$ NLSUSYGR. 
And the true vacuum of $N=2$ NLSUSY  model is achieved by the LSUSY model 
where all particles of th LSUSY supermultiplet are SUSY composites of NG fermions.   \par%
By the similar computations for (I) we obtain 
the vacuum which  breaks both SUSY and the local $U(1)$ spontaneously. 
\subsection{SU(5) Superon-quintet  model(SQM) of  particles}  
The graviton is the universal attractive force and dictates the evolution of 
$L_{SGM}$, which creates composite objects of all possible combinations of superons.  \\
We have found that the massless representation of sP  group is generated 
by the all possible parallel helicity products of supercharges. \\
And  the supercharge reduces to superon  in the low energy (current-field  identity in the low energy), i.e., the leading term of the supercurrent is the superon field in the low energy.   \\
The canonical quatization of superon is carried out in compatible with 
the SUSY algebra.  \\
NL/L SUSY  equivalence(relation) is shown for $N=2$ SUSY QED and 
$N=3$ SUSY YM, in $d=2$\cite{ST7}.  \\
%
%

%
From these considerations we speculate that all (observed ) particles are such composite objects of superons as eigenstates of sP group constructed group theoretically by the non-trivial multiplication products of the supercharge.  \\
SM, GUT may be regarded as the low energy effective theory of $N=10$ NLSUSYGR/SGM, where  ${N = {10} = {5}+{5^{*}}}$ under $SO(10) \supset SU(5)$, which we 
call {\it superon-quintet model(SQM)} of particles.
In order to survey the potential of SQM we adopt tentatively  the follwing naive L-R symmetric assignment 
for the SM particles:  \\
for $ (e, \nu_{e})$: $\delta^{ab}Q_{a}{Q^{*}}_{b}Q_{m}$, \ $(\mu, \nu_{\mu})$: $\delta^{ab}Q_{a}{Q^{*}}_{b}\epsilon^{lm}Q_{l}Q_{m}{Q^{*}_{n}}$,\ 
$(\tau, \nu_{\tau})$: 
$\epsilon^{abc}Q_{b}Q_{c}\epsilon^{ade}$${Q^{*}}_{d}{Q^{*}_{e}Q_{m}}$   
for $(u, d)$ : $\epsilon^{abc}Q_{b}Q_{c}Q_{m}$, \ \ 
$ (c, s)$ : $\epsilon^{lm}Q_{l}Q_{m}\epsilon^{abc}Q_{b}Q_{c}{Q^{*}}_{n}$, \ \
$(t, b)$ : $\epsilon^{abc}Q_{a}Q_{b}Q_{c}{Q^{*}}_{d}Q_{m}$,  \\ 
for the neutral Higgs particle we choose $Q_{a}{Q^{*}}_{a}Q_{l}{Q^{*}}_{l}$, 
\\   
for gauge bosons we have  \\
$Q_{a}{Q^{*}}_{a}, \ Q_{l}Q_{m}, \ Q_{l}{Q^{*}}_{m}, \ Q_{a}{Q^{*}}_{l}$,   \  $Q_{a}Q_{m}, \  \cdots$,   \\ 
$(a,b,\cdots=1,2,3; \ l,m, \cdots=4,5)$  \par
By specifying  the superon content of SM particles explicitly  we  interpret the Feynmann diagram of the SM/GUT 
in terms of the composite superon pictures, i.e. the single line of the propagator  of the SM particle in the Feynmann diagram is replaced by the multiple lines of the constituent.superons. 
We find that most Feynmann diagrams of the observed physical process of SM/GUT are reproduced in terms of the composite picture of particles. 
However the diagrams of the dangerous (no evidence) process in SM/GUT, e.g. FCNC and the proton decay, etc. are not reproduced due to the selection rurle for the superon number conservation at the vertex, i.e., 
the decay mode \ $p \rightarrow \bar{ e}+\nu+\pi^{0}$ \ is forbidden in SQM.  
The diagram of the dangerous process looks  forbidden  automatically 
in the tentative superon-quintet composite model for GUT.
As an example of the superon diagram,  $\beta$-decay diagram of SM is 
recasted as follows  in SQM.    \\[7mm]
%
%
%
%
%
\setlength{\unitlength}{1mm}
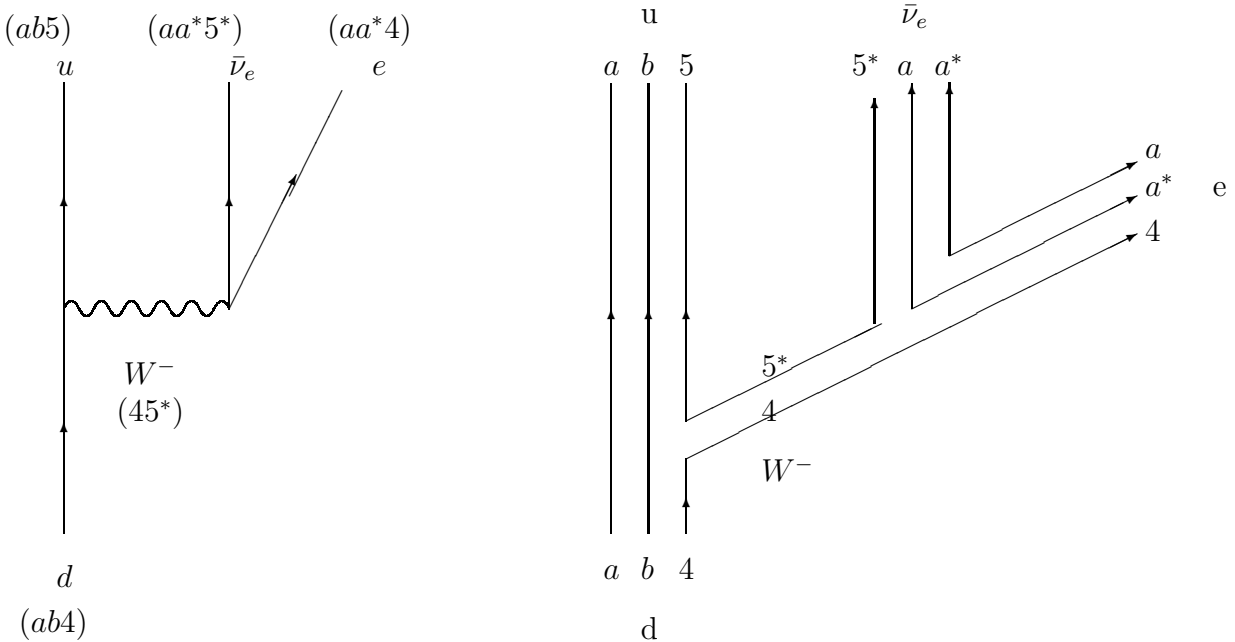
\begin{picture}(70,70)
\put(-1,-7){$d$}
\put(-6,-13){$(ab4)$}
\put(-8,66){$(ab5)$}
\put(-1,61){${u}$}
\put(0,0){\vector(0,1){15}}
\put(0,15){\vector(0,1){30}}
\put(0,45){\line(0,1){15}}
\put(22,30){\vector(0,1){15}}
\put(22,45){\line(0,1){15}}
\put(11,66){$(aa^{*}5^{*})$}
\put(22,61){${\bar{\nu}}_{e}$}
\put(22,30){\vector(1,2){8.9}}
\put(30,45){\line(1,2){7}}
\put(41,61){$e$}
\put(35,66){$(aa^{*}4)$}
\put(8,20){$W^{-}$}
\put(7,15){$(45^{*})$}

\qbezier(0,30)(1,32)(2,30)

\qbezier(2,30)(3,28)(4,30)

\qbezier(4,30)(5,32)(6,30)

\qbezier(6,30)(7,28)(8,30)

\qbezier(8,30)(9,32)(10,30)

\qbezier(10,30)(11,28)(12,30)

\qbezier(12,30)(13,32)(14,30)

\qbezier(14,30)(15,28)(16,30)

\qbezier(16,30)(17,32)(18,30)

\qbezier(18,30)(19,28)(20,30)

\qbezier(20,30)(21,32)(22,30)

\end{picture}
%
%
%
%
%
%
\setlength{\unitlength}{1mm}
\begin{picture}(70,70)
\put(0,0){\vector(0,1){30}}
\put(0,30){\line(0,1){30}}
\put(-1,-6){$a$}
\put(-1,61){$a$}
\put(5,0){\vector(0,1){30}}
\put(5,30){\line(0,1){30}}
\put(4,-6){$b$}
\put(4,61){$b$}
\put(4,68){u}
\put(9,-6){$4$}
\put(4,-14){d}
\put(10,0){\vector(0,1){5}}
\put(10,5){\line(0,1){5}}
\put(10,10){\vector(2,1){60}}


\put(9,61){$5$}
\put(10,15){\vector(0,1){15}}
\put(10.,30){\line(0,1){30}}
\put(10,15){\line(2,1){26}}

\put(35,28){\vector(0,1){30}}


\put(45,37){\vector(2,1){25}
}
\put(40,30){\vector(0,1){30}}

\put(40,30){\vector(2,1){30}}

\put(45,37){\vector(0,1){23}}

\put(20,15){$4$}
\put(20,21){$5^{*}$}
\put(32,61){$5^{*}$}

\put(38,61){$a$}
\put(43,61){$a^{*}$}
\put(80,45){e}
\put(37,68) {\bf{ ${\bar{\nu}}_{e}$}}
\put(20,7){$W^{-}$}
\put(71,50){$a$}
\put(71,45){$a^{*}$}
\put(71,39){$4$}
\end{picture} 
{} \\[18mm]

Revisiting SM and GUT from the diagramatic SQM composite viewpoints 
may give new insights into unsolved problems. 
%
%
As for the assignment of particles,  the different assignment $\epsilon^{abc}Q_{a}Q_{b}Q_{c}Q_{m}$ 
of the Higgs particle disfavors the left-right symmetric one and 
needs further study.
Finally we just mention that there is no (non-gravitational) excited states of quarks, leptons, gauge bosons, $\cdots$  in the SGM/SQM picture.  
\section{Cosmology of NLSUSYGR}
\subsection{Big Collapse of ultimate space-time(NLSUSYGR)}
Now we discuss the cosmological implications of NLSUSYGR. 
NLSUSYGR space-time described by  $L_{NLSUSYGR}$ is unstable and spontaneously collapses ({\it Big Collapse}) to $L_{SGM}$ of familiar  Riemann space-time {\it( graviton)}  and Nambu-Goldstone fermion matter {\it (superon)}. The Big Collapse may  induce the rapid expansion of three dimensional space-time  by the {\it  Pauli exclusion principle}\cite{TOL}. 
NLSUSYGR scenario  predicts the  {\it four}  dimensional space-time.  
Because we assume {\it space-time supersymmetry} at the local coordinate frame as the origin of ordinary  {\it particle supersymmetry},  i.e. we ask the isomorophism of $SO(1,D-1)$ and $SL(d,C)$:  
\begin{equation}
{D(D-1) \over 2}= 2(d^2 - 1), 
\end{equation} 
which holds  for  {\it only} 
\begin{equation}
D=4, \ \ d=2  
\end{equation}
and predicts  the ${\it four}$ dimensional space-time.  \\ 
The variation of SGM action  $L_{SGM}$ with respect to  ${e^{a}}_{\mu}$ gives 
the equation of motion for ${e^{a}}_{\mu}$ recasted as follows 
\begin{equation}
R_{\mu\nu}(e)-{1 \over 2}g_{\mu\nu}R(e)=
{8{\pi}G \over c^{4}} \{ \tilde T_{\mu\nu}(e,{\psi}^{j})-g_{\mu\nu}{c^{4}\Lambda \over 8{\pi}G} \}, 
\label{SGMEQ}
\end{equation}
where $\tilde T_{\mu\nu}(e,{\psi^{i}})$ $(i.j=1,2$ for $N=2$) 
abbreviates the stress-energy-momentum of superon(NG fermion) matter including the gravitational interactions. 
Note that $-{c^{4}\Lambda \over 8{\pi}G}$ can be interpreted as 
{\it the negative energy density of Riemann space-time}, 
i.e. {\it the dark energy density ${{\rho}_{D}}$}. 
(The negative sign is provided uniquely, which  produces simultaneously the correct kinetic term of superons in NLSUSY.) 
We anticipate that the graviton is the universal atrative force which dictates the evolution of the  soperon-graviton world (SGM action)  $L_{SGM}$ and constitutes gravitatonal (massless) conposites eigenststes of space-time symmetry $SO(N)$ sP, which ignites Big Bang model of the universe and continues to the SMs.
If superon-graviton phase(SGM action(\ref{L-exp})) with the cosmological constant  of space-time survives(swictched off)  in the evolution after Big Collapsec and oexists after the ignition of Big Bang, 
such SGM supace-tme  behaves as the dark side of the universe, 
i.e. the dark energy inducing the acceleration of the present universe, which is recognized only by the gravitational interaction. 
\subsection{Cosmology and particle physics}
On tangent flat space-time,  by the Noether theorem\cite{KS4} 
we obtain the conserved supercurrent 
${{S^{k\mu}}= i\sqrt{c^{4}\Lambda \over 8{\pi}G} \gamma^{a} \psi^k + \cdots},{} j, k=1,2$ \   associated with the NLSUSY invariance  
and obtains the  following superon(massless NG fermion)-vacuum coupling 
and $\sqrt{c^{4}\Lambda \over 8{\pi}G}$ is {\it the coupling constant $g_{sv}$ 
of superon with the vacuum via the supercurrent}. 
\begin{equation}
<  \psi^j_{\alpha} \vert {S^{k\mu}}_{\beta} \vert   0 > = 
i\sqrt{c^{4}\Lambda \over 8{\pi}G} (\gamma^{a})_{\alpha\beta} \delta^{jk},  
\label{LETH} 
\end{equation}
Further we have seen in the preceding section that the right hand side of (\ref{SGMEQ}) for N=2  
is essentially N=2 NLSUSY VA action  
and it is equivalent to the broken $N=2$ LSUSYQED action (\ref{L2action}) 
with the non-zero vacuum expectation value of the auxiliary field(Fayet-Iliopoolos term). \\
For the vacuum of the case (II) it gives the mass scale of the SUSY breaking  
\begin{equation}
{M_{SUSY}}^{2}  \sim \sqrt{c^{4}\Lambda \over 8{\pi}G}f\xi, 
\label{MASS1}
\end{equation}
to the component fields of the (massless) LSUSY supermultiplet. 
We find  NLSUSYGR(SGM) scenario gives interesting relations among the important quantities 
of the cosmology and the low energy particle physics, i.e., 
\begin{equation}
\rho_{D} \sim {c^{4}\Lambda \over 8{\pi}G}  \sim  g_{sv}{^2}. 
\label{MASS2}
\end{equation}
Suppose that in the (low energy) LSUSY supermultiplet 
the stable and the lightest massive particle retains the mass of the order of the spontaneous SUSY breaking.  \\
And if we identify the neutrino with such a particle and with $\lambda^{i}(x)$, 
i.e.
\begin{equation}
{{m_{\nu}}^{2} \sim \sqrt{c^{4}\Lambda \over 8{\pi}G}}, 
\label{numass}
\end{equation}
then SGM predicts the observed value of the (dark) energy density of the universe and 
naturally explains the mysterious numerical relations between $m_{\nu}$ and $\rho^{obs}_{D}$, \   provided $f\xi \sim O(1)$:
\begin{equation}
{\rho^{obs}_{D}} \sim (10^{-12}GeV)^{4} \sim {m_{\nu}}^{4}{} ( \sim  g_{sv}{^2} ).
\label{DENSITY}
\end{equation}
The tiny neutrino mass is the  direct evidence of SUSY (breaking) in the NLSUSYGR scenario, 
, i.e. Big collapse of space-time  and the subsequent creation of graitational (massless) composites 
in advance of the Big Bang.   
The extension to the large $N$ NLSUSY and the linearization of $\tilde T_{\mu\nu}(e,{\psi})$ necessary for building the realistic 
broken LSUSY model, 
which contains the mass scale in the higher order with $\psi$\cite{ST6}. \\
In Riemann flat space-time of SGM,  ordinary LSUSY gauge theory with the spontaneous SUSY breaking emerg as  (massless)  composites of NG fermion 
originating  from the NLSUSY cosmological constant of SGM.  
SM and GUT may be regarded as the low energy effective composite theory of NLSUSYGR in flat space-time.
\section{Summary}
%
We have proposed NLSUSYGR(SGM) scenario for unity of nature.  
The ultimate shape of nature is unstable ${\it four}$ dimensional space-time specified by $[x^{a}, {\psi_{\alpha}}^{N} ; x^{\mu}]$ and   described by NLSUSYGR $L_{\rm NLSUSYGR}( {w^{a}}_{\mu})$ equipping the cosmological term  with $\Lambda>0.$
Big Collapse(BC) of space-time  would occur
(due to false vacuum  $\Lambda >0$) 
and create ordinary Riemann space-time $[{ x^{a} ; x^{\mu}}]$ and massless fermionic matter {\it superon} ${\psi_{\alpha}}^{N}$, 
which is described  by SGM action $L_{\rm SGM}(e. \psi)$ .  \\
The universal attractive force graviton dictates the evolution of SGM world  
by forming all possible gravitational (massless) composites of superons 
which is, due to the field-current identity analogue,  equivalent to the (massless) representation of  sP algebra  
of space-time symmetry and ignites simultaneously 
the scenario of the Big Bang model of the Universe.  \par
The true vacuum configuration  
is achieved  by forming gravitational composite massless LSUSY supermultiplet and   
the subsequent oscillations around the true vacuum. 
We have shown in flat space-time that  broken $N$-LSUSY theory 
emerges from 
the NLSUSY cosmological term  of $L_{\rm SGM}(e,\psi)$ 
(NL/L SUSY relation). 
Interestingly SGM and its evolution  may regarded as the superfluidity of space-time and matter.
The vacua of composite SGM  scenario  
created by the BC of new space-time 
possesses  rich structures promising for the unified description of nature. 
SGM gives new  insights into the unsolved problems of cosmology and particle physics, for example, \  
the origin of the three-generations structure for quarks and leptons, \ the tiny neutrino mass, 
\  the proton decay, \ FCN, \ the dark matter and the dark energy, \ the space-time dimension {\it four}, \ etc.  
SGM predicts two color-singlet  particles, i.e., a  double-charge spin ${1 \over 2}$ fermion $E^{\pm 2}$: 
$\epsilon^{abc}Q_aQ_bQ_c\epsilon^{lm}{Q_{l}}^{*}Q_{ml}^{*}$ 
and  a neutral spin $1$  boson $S$:  $\delta^{ab}Q_aQ_b$, 
which may give clear signals in the high energy and the cosmic ray experiment. 
The decay modes of particles strongly depend upon the choice of the assignment of particles to the superon composites eigenstates 
of sP group.     \par
We have shown explicitly in 2 dimensional flat space-time  that $N=2$ LSUSY QED theory for the realistic 
$U(1)$ gauge theory  appears as the physical field configurations 
of the vacuum of $N=2$ NLSUSY theory on Minkowski space-time, which relate the particle physics with cosmology.    \par
In the NLSUSY/SGM scenario, the global SUSY and NG fermion superon 
play  essential roles  
and apparently the manifest local SUSY invariance may  be unnecessary 
in flat space so far as shown in the type (II) vacuum, 
for the massless NG fermion disappears from the physical states.   \par
Establishing  NL/L SUSY relation for  NLSUSYGR/SGM 
in curved space-time, i.e. linearizing directly SGM action $L_{SGM}(e,\psi)$ 
to find  the broken (global) LSUSY SUGRA-like equivalent theory 
with the mass generation in the tower of the helicity states of sP  
and the extension to large $N$, especially to ${N = {10} = {5}+{5^{*}}}$\cite{KS3}, is essential and under study.       \par
Finally we just mention that NLSUSYGR and the subsequent SGM scenario 
for the spin ${3 \over 2}$ NG fermion\cite{ST3, BAAK}  is in the same scope.  \\[10mm]
The author would like to thank Professor Y. Tanii,  Professor N. Yoshinaga,  Professor J. Sato and other colleagues of the theory group for the warm hospitality at Physics department of Saitama University.
\end{document}